\documentclass[aps,prb,twocolumn]{revtex4}

\usepackage{amsmath}
\usepackage{color}
\usepackage{amssymb}

\usepackage{graphicx}

\begin{document}

\title{ Influence of Exchange Scattering on Superfluid $^3$He states in Nematic Aerogel }

\author{V.P.Mineev$^{1,2}$}
\affiliation{$^1$Univ. Grenoble Alpes, CEA, INAC, PHELIQS, GT, F-38000 Grenoble, France\\
$^2$Landau Institute for Theoretical Physics, 142432 Chernogolovka, Russia}

\begin{abstract}
The superfluid state in bulk liquid $^3$He is realized  in the form of A or B phases.
Uniaxially anisotropic aerogel (nafen) stabilizes  transition from the normal to the polar superfluid state which on further cooling transitions  to the axipolar orbital glass state (Phys. Rev. Lett. {\bf 115}, 165304 (2015)). This is the case in nafen aerogel preplated by several atomic layers of $^4$He.
When   pure  liquid $^3$He fills the same nafen  aerogel a solid-like layer of $^3$He atoms coats the aerogel structure. 
The polar state is not formed anymore and a phase transition occurs directly to the axipolar phase (Phys. Rev. Lett. {\bf 120}, 075301 (2018)).
The substitution of $^4$He by $^3$He atoms  at the aerogel surface changes  the potential and adds the exchange scattering of quasiparticles on  the aerogel strands.
A calculation 
  shows that both of these effects
 can decrease  the degree of anisotropy of scattering and suppress the polar phase formation. 
The derived anisotropy of the spin diffusion
coefficient
in globally anisotropic aerogel is determined by the same parameter which controls the polar state emergence which  allows one to check  the effect of anisotropy change for different types of covering.

\end{abstract}

\date{\today}
\maketitle
\section{ Introduction.}
The superfluid state  of liquid $^3$He is formed by means the Cooper pairing with spin and orbital angular momentum equal to 1.
In isotropic space  the phase transition
depending on pressure
 occurs in either the A or  B superfluid phase \cite{Vollhardt2013}.
Investigation of superfluid phases in high porosity aerogel allows one to study the influence of impurities on superfluidity with nontrivial pairing \cite{Parpia1995,Halperin1995}.  It has been found that similar to  bulk $^3$He two superfluid A-like and B-like  phases exist in $^3$He in aerogel \cite{Osheroff2000}. However,  both the superfluid fraction and the temperature at which the superfluid is
manifested are suppressed  from their bulk values \cite{Parpia1995}. 
The interesting possibility is opened in globally anisotropic aerogel of lifting the degeneracy between the different superfluid phases with $p$-pairing. It was shown \cite{Aoyama2006} that in the case of  easy-axis anisotropy
 a new superfluid phase of $^3$He, the polar phase, is  stabilized below the transition temperature. It was also predicted \cite{Aoyama2006} that on further cooling a second-order transition into a polar-distorted A phase should occur. 
 Indeed, quite recently, the first observation of the polar phase was reported  \cite{Dmitriev2015}. This phase appears in $^3$He confined in new type of "nematically ordered" aerogel  called "nafen" with a nearly parallel arrangement of strands which play the role of ordered impurities. It was shown that in nafen
  the transition to the superfluid state always occurs to the polar phase and the region of its existence increases with density of strands.   
  In another type of nematically ordered but less dense and much less anisotropic  aerogel called "Obninsk aerogel" \cite{Dmitriev2015} the superfluid state  is always formed in the orbital  glassy A-like state.

To avoid a paramagnetic signal from surface solid $^3$He, the nafen samples  in the measurements  \cite{Dmitriev2015} were preplated by $\sim2.5$ $ ^4$He monolayers. 
Then the new  experiment series was performed with the same samples filled by
pure $^3$He \cite{Dmitriev2018}.
In this case the temperature of the superfluid transition is suppressed more strongly and  this effect increases with strands density such that in the most dense nafen the superfluid transition was not detected down to the lowest attained temperatures.
The superfluid transition occurs directly to the polar-distorted A-phase without the formation of an intermediate region of polar-state.  The small addition of $^3$He in the surface $^4$He layer, corresponding to 0.1 monolayer, also  completely kills superfluidity at 29,3 bar  in the most dense nafen, and in the less dense aerogel noticeably suppresses the critical temperature. In this case also, the transition occurs directly to the distorted A-state. Thus,  the situation looks as if the $^3$He covering  suppresses anisotropy necessary for the existence of the polar phase.

There was pointed out in Ref.7 : "The observed phenomena cannot
be explained by a change of the scattering specularity
because they are observed also at high pressures where
the scattering should be diffusive regardless of the
presence or absence of solid $^3$He". This statement is based on  previous studies (see references 5-9 in the paper \cite{Dmitriev2018}) 
of  the degree of specularity of $^3$He quasiparticles scattering on  metallic surfaces with different coverings. 
The corresponding information for liquid $^3$He filling nafen aerogel  is absent.
But, generally speaking, the substitution of $^4$He by $^3$He atoms  at aerogel surface changes  the potential and adds the exchange scattering of quasiparticles on  the aerogel strands.

I  study this problem taking into account  both the potential and the exchange scattering of quasiparticles of liquid $^3$He  on $^3$He atoms localized at the strands surface.
In Section II  I show  that $^3$He covering changes the intensity and the anisotropy of scattering.
  In the Section III the presented  derivation of spin diffusion current 
 shows that the anisotropy of the spin diffusion coefficient
in globally anisotropic aerogel is expressed through  the same  parameter  which determines the polar state emergence. 
Thus, being measured, the anisotropy decrease of spin diffusion in  nafen filled by  pure $^3$He  can serve as a direct indication of suppression of the temperature interval of  polar state existence.

\section{Superfluid $^3$He in uniaxially anisotropic aerogel with magnetic   scattering.}

The order parameter of superfluid phases of $^3$He is given \cite{Vollhardt2013} by the complex $3\times3$ matrix
$ A_{\alpha i} $, where $\alpha$  and $i$ are 
the indices  
numerating the Cooper pair wave function projections  on  spin and orbital axes respectively. All the phases with different order parameters $ A_{\alpha i} $ have the same critical temperature. The degeneracy is lifted
by the fourth-order terms  with respect to $ A_{\alpha i} $ in the Landau expansion of the free energy density. The most energetically profitable are 
the B-phase with the order parameter $A^B_{\alpha i}=\Delta R_{\alpha i}e^{i\varphi}$, where $R_{\alpha i}$ is a rotation matrix, and (in the high pressures region) the A-phase with the order parameter
\begin{equation}
A^A_{\alpha i}=\Delta V_\alpha(m_i+in_i),
\end{equation}
 where ${\bf V}$ is the unit spin vector and ${\bf m}$ and ${\bf n}$ are the orthogonal unit vectors  such that ${\bf m}\times${\bf n}=${\bf l}$ is the unit vector directed along the Cooper pair angular momentum.
 
The  different pairing states of superfluid $^3$He in a random medium 
with global uniaxial anisotropy can be compared  by making use  
the  second-order terms in the Landau free energy density. 
 They consist of an isotropic part, common to  all the  superfluid phases with 
 $p$-pairing, and the anisotropic part
\begin{eqnarray}
F^{(2)}=F^{(2)}_i+F^{(2)}_a~~~~~~~~~~~~~~~~~\nonumber\\=\alpha_0\left (\frac{T}{T_c}-1\right)A_{\alpha i}A^{\star}_{\alpha i}+\eta_{ij}A_{\alpha i}A^{\star}_{\alpha j},
\end{eqnarray}
where  $T_c=T_c( P )$ is the  transition temperature in the superfluid state suppressed with respect of transition temperature in the bulk liquid $T_{c0}( P )$. The medium  uniaxial anisotropy with anisotropy axis parallel to $\hat z$ direction coincident in our case with the average direction of aerogel strands is given by  the traceless tensor
 \begin{equation}
 \eta_{ij}=\eta\left (\begin{array}{ccc}1&0&0\\
 0&1&0\\
 0&0&-2\end{array}\right ).
 \label{eta}
 \end{equation}
 In the absence of global anisotropy $(\eta=0)$ all $p$-wave phases have the same critical temperature. 
 At positive $\eta>0$ the polar state with the order parameter of the form 
\begin{equation}
A_{\alpha i}=a V_\alpha z_i,
\label{polar}
\end{equation}
where $V_\alpha$ is the  unit spin vector,
has the lowest energy of anisotropy \cite{Mineev2013},
\begin{equation}
F_a=-2\eta|a|^2.
\label{a}
\end{equation}
Hence, it has the highest critical temperature $T_{c1}$ of transition from the normal state.
At some lower temperature $T_{c2}$ the polar state passes to the more energetically profitable
distorted A-state  \cite{Mineev2014} with the order parameter
 \begin{equation}
 A_{\alpha i}= V_\alpha\left [a\hat z_i+ib(\hat x_i\cos\varphi+\hat y_i\sin\varphi)\right ]
 \end{equation} 
 intermediate between the polar state at $b=0$ and the A-state at $a=b$.
 This state has 
the Cooper pair angular momentum $\hat{\bf l}=-\hat x \sin\varphi({\bf r})+\hat y \cos\varphi({\bf r})$  lying in the basal plane and locally ordered ($\varphi({\bf r})=const$) on the scales $L $ exceeding the coherence length $\xi_0$  but  smaller than the dipole length $\xi_d$  and randomly distributed on  scales larger than $L$.
The pure polar state exists in the temperature interval roughly determined by the energy of anisotropy difference between of the polar and the distorted A-states  \cite{Mineev2014},
\begin{equation}
T_{c1}-T_{c2}\approx \frac{\eta}{\alpha_0}T_c.
\end{equation}
Hence,  at small $\eta$ parameter  the temperature interval of the polar state existence is  small and  hardly observable.

The quasiparticle interaction with the nafen strands is modelled by the  interaction with the randomly distributed impurities including the globally anisotropic potential
and the globally anisotropic exchange part,
\begin{eqnarray}
H_{int}=\sum_i\int d^3r\psi^\dagger_\alpha({\bf r})\left [u({\bf r}-{\bf r}_i)\delta_{\alpha\beta}\right.\nonumber\\
\left.+J({\bf r}-{\bf r}_i)\mbox{\boldmath$\sigma$}_{\alpha\beta}{\bf S}
\right ]\psi_\beta({\bf r}),
\label{int}
\end{eqnarray}
where  $ {\bf S}$ is the spin of the impurity and $\mbox{\boldmath$\sigma$}$ are the $^3$He quasiparticles spin matrices.
The exchange scattering in an isotropic aerogel has been considered by Sauls and Sharma \cite{Sauls2003} and by Baramidze and Kharadze \cite{Baramidze2003}.
They  have shown that 
if the scattering amplitude on impurities includes an exchange part
then the critical temperatures splitting of  $A_1$ and $A_2$ transitions under an external field H
decreases in comparison with  the impurity free case:
$$
T_{A_1}-T_{A_2}=(\gamma_0-\gamma_{imp})H.
$$
The effect arises due to  an interference between  scalar and exchange scattering such that
$$
\gamma_{imp}\propto uJ
$$
 is proportional to the product of the corresponding amplitudes.
In the NMR experiments \cite{Dmitriev2015,Dmitriev2018}  the field is small and this effect is negligible, but one needs to consider an influence of the globally anisotropic scattering on critical temperature.

To find the critical temperature of superfluid transition
in globally anisotropic aerogel one must 
calculate the second-order terms in the Landau free energy density
\begin{widetext}
\begin{equation}
F^{(2)}=\frac{1}{3}\left \{\frac{1}{g}\delta_{ij}- T\sum_{\omega}\int\frac{d^3p}{(2\pi)^3}\hat p_i\Gamma_j({\bf p},\omega,)G({\bf p},\omega)G(-{\bf p},-\omega)
\right\}A_{\mu i}^\star A_{\mu j},
\label{fe}
\end{equation}
\end{widetext}
where $g$ is the constant of $p$-wave triplet pairing.
Here,\begin{equation}
G({\bf p},\omega)=\frac{1}{i\omega-\xi_{\bf p}-\Sigma_{\bf p}(\omega)}
\label{G}
\end{equation}
 is the normal state quasiparticle Green function and $\Gamma_j^{\mu\nu}({\bf p},\omega)$ is the vertex part. $\xi_{\bf p}=\varepsilon_{\bf p}-\mu$ is the quasiparticles energy  counted from the chemical potential, and $\omega=\pi T(2n+1)$ is the fermion Matsubara frequency.  The Planck constant $\hbar$ was everywhere put equal to 1.
The self-energy part is given by the equation
\begin{equation}
\Sigma_{\bf p}(\omega)=\int\frac{d^3p^\prime}{(2\pi)^3}{U_{{\bf p}-{\bf p}^\prime}^2}G({\bf p}^\prime,\omega).
\label{sigma}
\end{equation}
Here, according to Abrikosov and Gor'kov \cite{Abrikosov1961}, the "impurity line" $U_{{\bf p}-{\bf p}^\prime}^2$ arises after averaging over impurity positions and also over the orientation of the spins of all impurity atoms,
$\langle S_iS_k\rangle=\frac{1}{3}S(S+1)\delta_{ik}$, where in our particular case $S=1/2$. Then taking into account $\sigma_{\alpha\gamma}^i\sigma_{\gamma\alpha}^i=\frac{3}{4}$ we obtain
\begin{eqnarray}
U_{\bf p}^2=n_i\left [u_{\bf p}^2+ \langle{S_i\sigma_{\alpha\gamma}^iS_k\sigma_{\gamma\alpha}^k}\rangle J_{\bf p}^2\right]\nonumber\\
=n_i\left[u_{\bf p}^2+\frac{1}{4}S(S+1)J_{\bf p}^2\right]
,
\label{U}
\end{eqnarray}
where $n_i$ is impurity concentration and $u({\bf p})$ and  $J({\bf p})$ are the Fourier transforms of the amplitudes of potential and exchange scattering from Eq.(\ref{int}).
According to assumption about global anisotropy
they  depend on the momentum direction such that
\begin{equation}
n_iu_{\bf p}^2=\frac{1}{2\pi N_0\tau_p}\left \{1-\delta_p\left [\hat p_z^2-\frac{1}{2}(\hat p_x^2+\hat p_y^2)\right ]\right \},
\label{p}
\end{equation}
\begin{equation}
n_iJ_{\bf p}^2=\frac{1}{2\pi N_0\tau_{ex}}\left \{1-\delta_{ex}\left [\hat p_z^2-\frac{1}{2}(\hat p_x^2+\hat p_y^2)\right ]\right \},
\label{e}
\end{equation}
where
$N_0$ is the density of states per one spin projection,  $\hat p_i$ are  the projections of  momentum unit vector $\frac{{\bf p}}{|{\bf p}|}$ on the $i=(x,y,z)$ coordinate axis, $\tau_p$, and $\tau_{ex}$ are the isotropic parts of mean free time of potential and exchange scattering and $\delta_p$ and $\delta_{ex}$ are 
the corresponding degree of anisotropy. The anisotropic part of $U_{\bf p}^2$ is taken with the sign opposite to that in Ref.5 and chosen such that $\int\frac{d\Omega}{4\pi}U_{\bf p}^2$ is independent of the anisotropic part of scattering.

So, the self energy obtained from Eqs.(\ref{sigma})-(\ref{e}) is 
 \begin{equation}
\Sigma_{\bf p}(\omega)=-\frac{i}{2\tau}\left \{1-\delta\left [\hat p_z^2-\frac{1}{2}(\hat p_x^2+\hat p_y^2)\right ]\right \}
{\text sign}~\omega.
\label{sigma1}
\end{equation}
Along with the isotropic term it includes a term describing the
global uniaxial anisotropy. 
Each of these terms consists  of  two independent parts: the potential part  and the exchange one determined in the following way
\begin{equation}
\frac{1}{\tau}=\frac{1}{\tau_p}+\frac{1}{\tau_{ex}},~~~~~~~ \frac{\delta}{\tau}=\frac{\delta_p}{\tau_p}+\frac{\delta_{ex}}{\tau_{ex}}.
\label{anis}
\end{equation} 

The vertex part must be found from the integral equation
\begin{widetext}
\begin{equation}
\Gamma_j({\bf p},\omega)=\hat p_j+n\int\frac{d^3p^\prime}{(2\pi)^3}\left [u_{{\bf p}-{\bf p}^\prime}^2 
+\frac{1}{3}S(S+1)(g^\dagger)^\mu_{\alpha\beta}\sigma^i_{\lambda\alpha}\sigma^i_{\rho\beta}g^\mu_{\lambda\rho} J^2_{{\bf p}-{\bf p}^\prime}
\right ]\Gamma_j({\bf p}^\prime,\omega)G({\bf p}^\prime,\omega)G(-{\bf p}^\prime,-\omega).
\label{Gamma}
\end{equation}
\end{widetext}
It  is known \cite{Abrikosov1961} that for the case of singlet superconductivity  the exchange part of scattering in this equation is given by
\begin{equation}
\frac{1}{3}S(S+1) g^t_{\alpha\beta}\sigma^i_{\lambda\alpha}\sigma^i_{\rho\beta}g_{\lambda\rho}J^2_{\bf q}=-\frac{1}{4}S(S+1)J^2_{\bf q},
\end{equation}
where the matrix $\hat g=\left (\begin{array}{cc}0&1\\-1&0\end{array}\right )$, and the superscript $t$ indicates transposition.
As result, there are two different "scattering time" originating from the self-energy  and the vertex  \cite{Abrikosov1961}.
The corresponding combination for the triplet pairing is
\begin{equation}
\frac{1}{3}S(S+1) (g^\dagger)^\mu_{\alpha\beta}\sigma^i_{\lambda\alpha}\sigma^i_{\rho\beta}g^\nu_{\lambda\rho}J^2_{\bf q}=\frac{1}{4}S(S+1)J^2_{\bf q}\delta_{\mu\nu},
\end{equation}
where
$g^\nu_{\lambda\rho}=(-\sigma^z_{\lambda\rho},i\delta_{\lambda\rho},\sigma^x_{\lambda\rho})$,
such that the "scattering time"  originating from the self-energy  and the vertex are equal to each other. 
Thus, the Eq.(\ref{Gamma}) is
\begin{eqnarray}
\Gamma_j({\bf p},\omega)=\hat p_j~~~~~~~~~~~~~~~~~~~~~~~~\nonumber\\+\int\frac{d^3p^\prime}{(2\pi)^3}
U_{{\bf p}-{\bf p}'}^2\Gamma_j({\bf p}^\prime,\omega)G({\bf p}^\prime,\omega)G(-{\bf p}^\prime,-\omega)
\label{Gamma'}
\end{eqnarray}
and its  solution has the form
\begin{equation}
\Gamma_j^{\mu\nu}(\omega,{\bf p})=\left \{\hat p_j+\Gamma_\omega\left [\hat p_z\hat z_j-\frac{1}{2}(\hat p_x\hat x_j+\hat p_y\hat y_j)\right ]
  \right\},
\end{equation}
where for the $\delta\ll1$ 
\begin{equation}
\Gamma_\omega=\frac{\delta}{3\tau\left |\omega+\frac{1}{2\tau}{\text sign}~\omega\right |} +{\cal O}(\delta ^2).
\end{equation}
Substitution of  the vertex $\Gamma_j^{\mu\nu}(\omega,{\bf p})$ and the Green function $G({\bf p},\omega)$ into Eq.(\ref{fe}) yields
\begin{eqnarray}
F_2=\alpha A_{\mu i}^\star A_{\mu i}-2\eta \left [A_{\mu z}^\star A_{\mu z}-\frac{1}{2}(A_{\mu x}^\star A_{\mu x}+A_{\mu y}^\star A_{\mu y})\right ],
\label{main}
\end{eqnarray}
where
\begin{eqnarray}
\alpha=\frac{N_0}{3}\left [ \ln\frac{T}{T_{c0}}+\psi\left ( \frac{1}{2}+\frac{1}{4\pi T\tau}\right ) -\psi\left ( \frac{1}{2} \right )\right.\nonumber\\ 
\left.-\frac{1}{5}\frac{\delta}{4\pi T\tau}
\psi^{(1)}\left ( \frac{1}{2}+\frac{1}{4\pi T\tau}\right ) \right ],~~~~~~~~~~~~
\end{eqnarray}
\begin{equation}
\eta=\frac{8N_0}{45}\frac{\delta}{4\pi  T\tau}\psi^{(1)}\left ( \frac{1}{2}+\frac{1}{4\pi T\tau}\right ).
\end{equation}
Here, $\psi(z)$, and $\psi^{(1)}(z)$ are the digamma function and its first derivative.

At $\delta>0$  the critical temperature of the phase transition to the superfluid state is maximal for the polar phase Eq.(\ref{polar}),
and is determined by the equation
\begin{equation}
\alpha -2\eta=0.
\label{eq}
\end{equation}
In the  limit of weak scattering
$ 4\pi T\tau_c>>1$
the transition to the polar state occurs at
\begin{equation}
T_{c1}=T_{co}-\frac{\pi}{8\tau}+\frac{11\pi }{60\tau}\delta.
\end{equation}
It is worthwhile to recall that  at small degrees of anisotropy  the temperature interval of the polar state existence is  small and  hardly observable.

According to Eq.(\ref{anis}) the degree of global anisotropy $\delta$ is determined by two independent terms originating from the potential and the exchange scattering. The latter can in principle suppress the anisotropy. 
However, the  quasiparticle self-energy and the vertex part  due to exchange scattering have the same structure 
as for pure potential scattering.
Hence,  the anisotropy suppression can  also originate from the potential scattering which is different for the covering of aerogel stands by a solid $^3$He layer.

The change in anisotropy of scattering  for different types of covering   must also reveal itself in the changes of spin diffusion anisotropy.
In the next section
I derive
the normal liquid $^3$He spin diffusion current  flowing through the media filled by the randomly distributed impurities with globally anisotropic potential  and exchange scattering.

\section{Spin current}

The spin current in neutral Fermi liquid is calculated  \cite{,Makhlin1992,Volovik1992} as the response to  the gradient of angle of rotation of the spin space 
$\mbox{\boldmath$\omega$}_i=\nabla_i\mbox{\boldmath$\theta$}$,
\begin{equation}
{\bf j}_{i}=-\frac{\delta H }{\delta\mbox{\boldmath$\omega$}_i},
\end{equation}
where
\begin{equation}
H=\frac{1}{2m}\int d^3r (D_i^{\alpha\lambda}\psi_\lambda)^\dagger D_i^{\alpha\mu}
\psi_\mu+H_{int},
\end{equation}
\begin{equation}
D_i^{\alpha\beta}=-i\delta_{\alpha\beta}\nabla_i+\frac{1}{2}\mbox{\boldmath$\sigma$}_{\alpha\beta}\mbox{\boldmath$\omega$}_i,
\end{equation}
and $H_{int}$ includes the Fermi liquid interaction and  the interaction with  impurities, Eq.(\ref{int}).

 At low temperatures  the collisions between the Fermi liquid quasiparticles induce negligibly small correction to the spin diffusion due to the scattering on aerogel strands. On the other hand,
  we are mainly interested in the spin current anisotropy in the anisotropic media and will ignore the temperature dependence of exchange amplitude of scattering due to the Kondo effect \cite{Edelstein1983}. In this case one can work with the field theory technique for T=0. The response of the gauge field $\mbox{\boldmath$\omega$}_i$ is calculated in complete analogy with response to the usual vector potential $A_i$ in the calculation 
 of electric current in an isotropic metal with randomly distributed impurities performed in \cite{Abrikosov1959}. The spin current  at finite wave vector ${\bf k}$ and external frequency $\omega$ is
 \begin{eqnarray}
 {\bf j}_{i}({\bf k},\omega)=~~~~~~~~~~~~~~~~~~~~~~
 \nonumber\\\frac{i}{4m}Tr\int_{-\infty}^{+\infty}\frac{d\varepsilon}{2\pi}\int\frac{d^3p}{(2\pi)^3}p_i\mbox{\boldmath$\sigma$}_{\alpha\beta}
(\mbox{\boldmath$\sigma$}_{\beta\alpha}\mbox{\boldmath$\omega$}_j)\Pi_j-\frac{1}{4}n\mbox{\boldmath$\omega$}_i,
\label{spcur}
 \end{eqnarray}
where 
 $n$ is the number of liquid $^3$He atoms in the unit volume, function $\Pi_j$ is determined by the equation
\begin{eqnarray}
\Pi_j(p,p-k)=G({\bf p},\varepsilon)G({\bf p}-{\bf k},\varepsilon-\omega)\nonumber\\
\times\left [p_j+\int\frac{d^3p'}{(2\pi)^3}U^2({\bf p}-{\bf p}')\Pi_j(p',p'-k)\right ],
\label{pi}
\end{eqnarray}
$p=({\bf p},\varepsilon)$, $k=({\bf k},\omega)$,
\begin{equation}
G({\bf p},\varepsilon)=\frac{1}{\varepsilon-\xi_{\bf p}-\Sigma_{\bf p}(\varepsilon)},
\end{equation}
$U^2({\bf p})$ is determined by Eq.(\ref{U}),  and $\Sigma_{\bf p}(\varepsilon)$ is given by Eq.(\ref{sigma1}). The vertex correction does not introduce changes in the spin structure of Eq.(\ref{spcur}) due to the identity $\sigma^i_{\alpha\lambda}\sigma^p_{\lambda\mu}\sigma^p_{\rho\alpha}\sigma^j_{\mu\rho}=\sigma^i_{\alpha\beta}\sigma^j_{\beta\alpha}$. 

At ${\bf k}=0,~\omega=0$ the first term in the current expression (\ref{spcur}) cancels out the second "diamagnetic" term. We are interested in calculating the current at ${\bf k}=0,\omega\ne0$. In this case,
\begin{equation}
 {\bf j}_{i}=\frac{i}{4m}Tr\int_{0}^{\omega}\frac{d\varepsilon}{2\pi}\int\frac{d^3p}{(2\pi)^3}p_i\mbox{\boldmath$\sigma$}_{\alpha\beta}
(\mbox{\boldmath$\sigma$}_{\beta\alpha}\mbox{\boldmath$\omega$}_j)\Pi_j({\bf k}=0)
\label{currant}
\end{equation}
and
the solution of Eq.(\ref{pi}) in a linear approximation with respect to $\delta$ and at $\omega\tau\ll1$ is
\begin{eqnarray}
\Pi_j({\bf k}=0)=G({\bf p},\varepsilon)G({\bf p},\varepsilon-\omega)
\nonumber\\
\times\left \{p_j+\frac{2}{3}\delta\left [\hat p_z\hat z_j-\frac{1}{2}(\hat p_x\hat x_j+\hat p_y\hat y_j)\right ]\right\}.
\end{eqnarray}
Substituting this into Eq.(\ref{currant}), we obtain
\begin{widetext}
\begin{equation}
 {\bf j}_{i}=\frac{1}{6}\left\{\delta_{ij}+\frac{16}{15}\delta\left [\hat z_i\hat z_j-\frac{1}{2}(\hat x_i\hat x_j+\hat y_i\hat y_j) \right ]\right\}i\omega\tau N_0 v_F^2
 \mbox{\boldmath$\omega$}_j.
 \end{equation}
 Here $v_F$ is the Fermi velocity. 
 Making use of the Larmor theorem 
\begin{equation}
\gamma{\bf H}=\frac{\partial\mbox{\boldmath$\theta$}}{\partial t}=-i\omega\mbox{\boldmath$\theta$},
\end{equation}
where $\gamma=2\mu$ is the gyromagnetic ratio, and $\mu$ is the magnetic moment of $^3$He atoms, one can rewrite the expression for current as
\begin{equation}
 {\bf j}_{i}=-\frac{1}{3}\left\{\delta_{ij}+\frac{16}{15}\delta\left [\hat z_i\hat z_j-\frac{1}{2}(\hat x_i\hat x_j+\hat y_i\hat y_j) \right ]\right\}
 \tau N_0 v_F^2\mu\nabla_j{\bf H}.
 \end{equation}
 To rewrite the spin current as the magnetic diffusion current one should multiply both sides of this equation by $2\mu$ to obtain
 \begin{equation}
  {\bf j}^M_{i}=-\frac{1}{3}\left\{\delta_{ij}+\frac{16}{15}\delta\left [\hat z_i\hat z_j-\frac{1}{2}(\hat x_i\hat x_j+\hat y_i\hat y_j) \right ]\right\}
 \tau  v_F^2\nabla_j{\bf M},
 \end{equation}
 \end{widetext}
where the Fermi-liquid magnetization is ${\bf M}=2\mu^2N_0{\bf H}$. Thus, the spin diffusion currents along the direction of nafen strands  and in the direction perpendicular to them are
\begin{eqnarray}
{\bf j}^M_{z}=-\frac{1}{3}\left\{1+\frac{16}{15}\delta
 \right\} \tau  v_F^2\nabla_z{\bf M},\\
  {\bf j}^M_{x}=-\frac{1}{3}\left\{1-\frac{8}{15}\delta
  \right\} \tau  v_F^2\nabla_x{\bf M}
\end{eqnarray}
respectively. One can demonstrate that a similar calculation taking into account the Fermi liquid renormalization adds in these formulas the pre-factor $(1+F_0^a)(1+F_1^a/3)$.

Thus, the anisotropy of the spin diffusion coefficient  is expressed through the same parameter of anisotropy $\delta$ that determines the temperature interval of the polar state existence.

 \section{Conclusion}

It was shown that the  degree of global anisotropy  responsible for  polar state stability  is determined by two  mechanisms,  originating from the potential and the exchange scattering. The suppression of anisotropy  narrows the temperature interval of the polar state existence, making it hardly observable. The anisotropy decrease can be controlled by the measurements of spin diffusion  because  
the difference in the spin diffusion coefficients in directions parallel and perpendicular to nafen strands is found to be proportional to the same parameter that determines the polar state emergence.

The authors of the paper \cite{Dmitriev2018} have pointed out the dominate role of the exchange scattering  in the anisotropy suppression (see the citation  of Ref.7 in the Introduction). However, according to the presented results,
the exchange interaction yields  the quasiparticle self-energy and the vertex part  of the same structure as for pure potential scattering.
Hence, although the exchange mechanism works only in the case of coating  of aerogel strands by a solid $^3$He layer it is possible that the main role in the anisotropy
decrease is played  by  the change of potential scattering with aerogel strands covered by $^3$He instead of $^4$He. 
The problem  of choosing between  the two
mechanisms of anisotropy suppression
will be addressed in future investigations.

 Being mainly interested in the role of anisotropy of exchange scattering, I neglect  throughout this  paper   the possible temperature dependence of the amplitude of exchange scattering due to the Kondo effect.  The logarithmic increase of positive ion mobility starting at T=50 mK up to the superfluid transition temperature ( see the paper \cite{Edelstein1983} and references therein)    means that the exchange scattering has a ferromagnetic character, in agreement with the notion that $^3$He is an almost ferromagnetic Fermi liquid. Thus, the Kondo effect  weakens  the magnitude of the pair breaking by magnetic scattering.

\acknowledgments

I am indebted to  V.Dmitriev for helpful and stimulating discussions.

\end{document}